\documentclass[twocolumn,showpacs,preprintnumbers,amsmath,amssymb,prb]{revtex4-1}
\usepackage{graphicx}
\usepackage{amsmath, amsthm, amssymb}
\usepackage{latexsym}
\usepackage{amssymb}
\usepackage{bm}
\usepackage{dcolumn}
\usepackage{epstopdf}
\usepackage{color}

\usepackage{natbib}
\usepackage[normalem]{ulem}

\begin{document}

\title{Magnetic Field Induced Vortex Lattice Transition in HgBa$_{2}$CuO$_{4+\delta}$}

\author{Jeongseop A. Lee$^{1}$, Yizhou Xin$^{1}$, I. Stolt$^{1}$, W. P. Halperin$^{1}$, A. P. Reyes$^{2}$ P. L. Kuhns$^{2}$, M. K. Chan$^{3}$}

\affiliation{$^1$Department of Physics and Astronomy Northwestern University, Evanston, Illinois 60208, USA\\$^2$National High Magnetic Field Laboratory, Tallahassee, Florida 32310, USA\\$^3$Pulsed Field Facility, National High Magnetic Field Laboratory, Los Alamos National Laboratory, Los Alamos, New Mexico 87545, USA}
\date{Version \today}

\begin{abstract}

Measurements of the $^{17}$O nuclear magnetic resonance (NMR)  quadrupolar spectrum of  apical oxygen in HgBa$_{2}$CuO$_{4+\delta}$  were performed over a range of magnetic fields from 6.4 to 30\,T in the superconducting state.  Oxygen isotope exchanged single  crystals were investigated with doping corresponding to superconducting transition temperatures from 74\,K underdoped, to 78\,K overdoped.  The apical oxygen site was chosen since its NMR spectrum has narrow quadrupolar satellites that are well separated from any other resonance. Non-vortex contributions to the spectra can be deconvolved in the time domain to determine the local magnetic field distribution from the vortices.  Numerical analysis using Brandt's Ginzburg-Landau theory was used to find structural parameters of the vortex lattice, penetration depth, and coherence length as a function of magnetic field in the vortex solid phase. From this analysis we report a vortex structural transition near 15\,T from an oblique lattice with an opening angle of $73^{\circ}$ at low magnetic fields to a triangular lattice with $60^{\circ}$ stabilized at high field.  The temperature for onset of vortex dynamics has been identified with vortex lattice melting. This is independent of the magnetic field at sufficiently high magnetic field similar to that reported for YBa$_2$Cu$_3$O$_7$ and Bi$_{2}$Sr$_{2}$CaCu$_{2}$O$_{8+\delta}$ and is  correlated with mass anisotropy of the material.  This behavior is accounted for theoretically only in the limit of very high anisotropy.\end{abstract} 

\pacs{ }

\maketitle

\section{Introduction}

The vortex state of high temperature superconductors (HTS) dominates the magnetic field-temperature phase diagram at  fields between a very large upper critical field, $H_{c2}$, and an extremely small lower critical field, $H_{c1}$.  Consequently, vortices play an essential role in the behavior of cuprate superconductors.  Technological applications of HTS depend on understanding and controlling the vortex dynamics.  Driven by the Lorentz force from an electrical current, they dissipate heat and adversely affect the superconducting critical current.  The vortex state of the ideal defect-free material forms a lattice as was predicted by Abrikosov for type II superconductors~\cite{Abrikosov1957}.  The  physical structure of this vortex lattice (VL)  in HTS reflects the underlying mass anisotropy of the crystal and the nature of the supercurrents circulating around the vortex core having a radius the size of the coherence length, a few nm in size.  At high magnetic field the supercurrents  from nearby vortices strongly overlap with one another, which together with the electronic structure at the vortex core, determine the symmetry of the VL as well as the vortex dynamics that result from  thermal fluctuations or external forces.  The effect of vortex-vortex interactions is most easily investigated by varying the vortex density which is proportional to the applied  field.  In the present work over the range of fields from  $H = 6.4$ to 30\,T, we have used $^{17}$O nuclear magnetic resonance (NMR)~\cite{Mounce2012}, finding a vortex-lattice transition in the simple, single-layer, cuprate superconductor HgBa$_{2}$CuO$_{4+\delta}$ (Hg1201).

Thermal fluctuations of vortices are particularly evident in highly anisotropic, layered superconductors, such as HTS cuprates~\cite{Nelson1988,Brandt1989}.  As the temperature is decreased  below $T_{c}$ in a defect free crystal, the vortices undergo a sharp thermodynamic first-order phase transition into a pinned vortex solid phase.  For example, this was demonstrated by  Kwok {\it et al.}~\cite{Kwok1994} who observed the corresponding magnetization discontinuity in YBa$_2$Cu$_3$O$_7$ (Y123) untwinned crystals. The condensation of vortices into a lattice at a melting transition, $T_m$,  is analogous to the first order liquid-solid phase transition of matter, phenomenologically  accounted for by the Lindemann criterion for melting~\cite{Brandt1989,Glazman1991}.

At the vortex melting transition there is an abrupt change in the frequency of the dynamics.  In the vortex liquid phase the spatial fluctuations of the local magnetic field from vortex supercurrents are motionally averaged to zero on the time scale of an NMR measurement, $t \ll 10$\,ns, given by the Larmor period.  However, in the vortex solid phase the local field distribution is quasi-static with a relevant time scale, $t > 0.1$\,ms, and the internal vortex field profile contributes to the NMR spectrum which can be used to determine the VL structure as well as the superconducting penetration depth and coherence length. This information is similar to what can be inferred from muon spin resonance, although it is a technique generally restricted to relatively low magnetic fields.  

A rich variety of magnetic and electronic orders are evident in the low field vortex state including unusual checkerboard type electronic inhomogeneity as reported for Bi$_{2}$Sr$_{2}$CaCu$_{2}$O$_{8+\delta}$ (Bi2212)~\cite{Hoffman2002, Wise2008, Mounce2011}. However,  in very high magnetic fields,   where vortex-vortex interactions become particularly strong, this behavior is less well-established.  Exploring the high field region is the main thrust of our work where we present  $^{17}$O NMR measurements of vortex structure and dynamics in Hg1201 single crystals at magnetic fields as high as  $H = 30$\,T. 

\section{Experimental Details}

Crystals of Hg1201 were  grown at the University of Minnesota.  Isotope exchange for $^{17}$O  NMR was  performed at Northwestern University followed by annealing for typically one week  to establish the necessary doping and homogeneity.  Characterization was performed using $^{17}$O and $^{199}$Hg  nuclear magnetic resonance (NMR), Laue x-ray diffraction, and low field SQUID measurements, the latter in order to determine the superconducting $T_{c}$.  The suite of single crystal samples of Hg1201 include those underdoped with $T_{c}$ = 74\,K (UD74), 79\,K (UD79), 87\,K (UD87),  optimally doped with $T_{c}$ = 94\,K (OP94) and overdoped 78\,K (OD78), where some of these samples have been studied previously~\cite{Mounce2013,Lee2016}.  Their  oxygen concentrations correspond  to hole doping of $p =  0.095$,  0.105, 0.118, 0.130, and 0.216  respectively, obtained by comparing the measured $T_{c}$ with the phase diagram, Fig.\ref{fig1}~\cite{Yamamoto2000}.  Transition widths from magnetization measurements for all the samples were less than 5\,K.  NMR measurements were taken with the applied magnetic field  aligned with the crystal $c$-axis where there exists two sets of five $^{17}$O NMR spectral peaks: one set for oxygen O(1) in the CuO$_2$ plane, and another for the apical oxygen, O(2).  Since the nuclear spin is $I=5/2$, these are associated with the central transition, ($\frac{1}{2}$,-$\frac{1}{2}$), and four quadrupolar satellites corresponding to the transitions, ($\frac{5}{2}$,$\frac{3}{2}$), ($\frac{3}{2}$,$\frac{1}{2}$), (-$\frac{1}{2}$,-$\frac{3}{2}$), and (-$\frac{3}{2}$,-$\frac{5}{2}$). In general the transition frequencies are labeled by indices ($m, m-1$) where $m$ is the quantum number for the energy levels.  For this investigation we have varied the magnetic field from $H$ = 6.4 to 30\,T over a range of temperature, from $\sim4$\,K to 100\,K.  NMR measurements with magnetic fields of 14\,T and below were performed at Northwestern University.  For magnetic fields above 14\,T the measurements were performed at the National High Magnetic Field Laboratory (NHMFL) in cell 2 of the DC-field facility.  

$^{199}$Hg NMR was used to verify  doping concentrations since the dopant oxygen resides in the HgO plane and  its concentration can be correlated with secondary NMR peaks in the spectrum~\cite{Haase2012, Mounce2013, Lee2016}.

\section{Vortex Lattice Melting}

\begin{figure}[!ht]
\centerline{\includegraphics[width=0.5\textwidth]{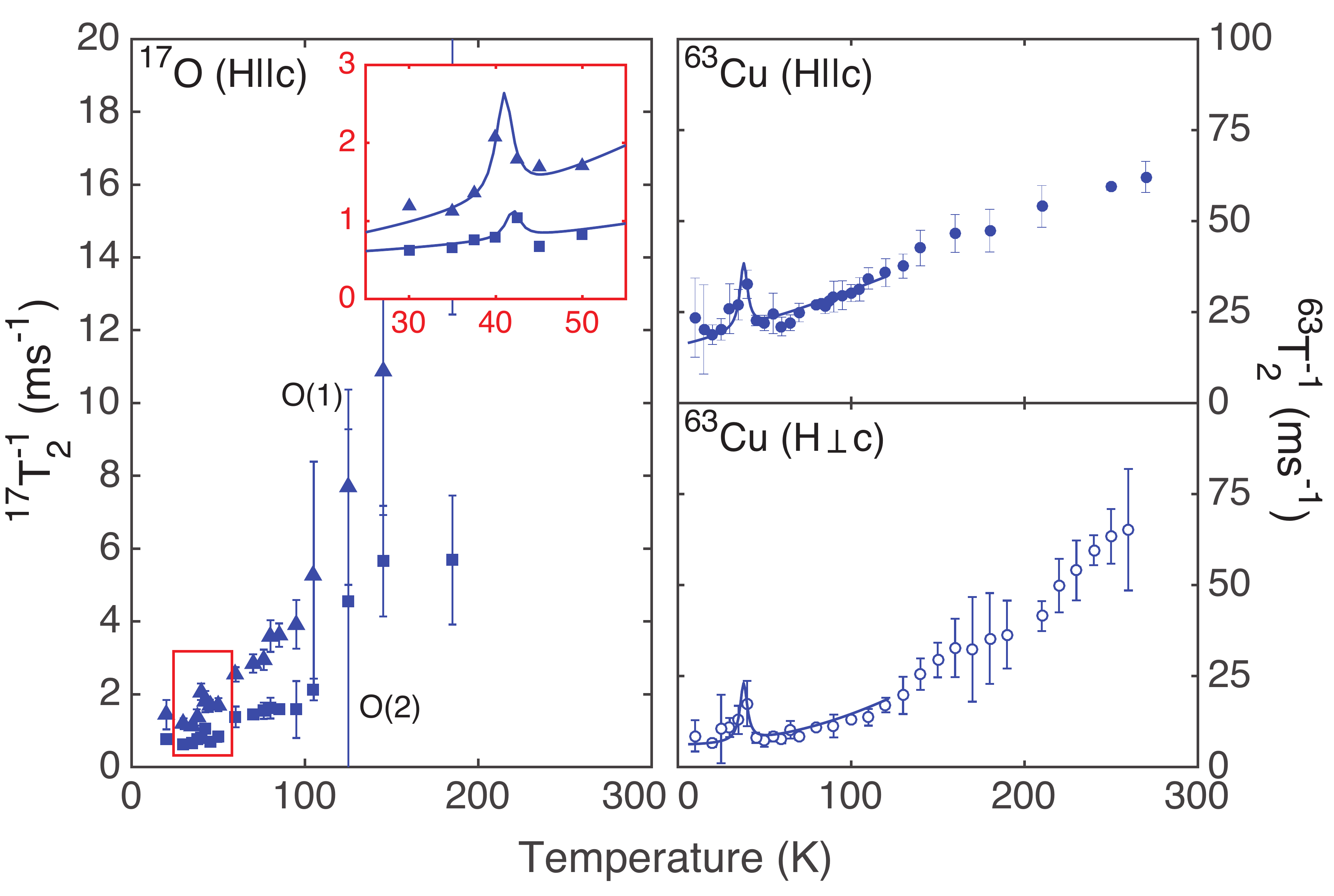}}
\caption{The spin-spin relaxation rate for $^{17}$O  and $^{63}$Cu NMR showing the vortex melting transition in  sample UD87 for $H= 6.4$\,T.  This transition is indicated by a peak in the relaxation rate when  the magnetization decay is analyzed as a pure gaussian~\cite{Bachman1998}. More precisely, the appearance of a solid vortex on cooling is marked by the onset of a lorentzian component to the otherwise gaussian recovery~\cite{Bachman1998}.}
\label{fig1}
\end{figure}

The spin-spin relaxation rates, $T_{2}^{-1}$, of $^{63}$Cu and $^{17}$O in the CuO$_2$ plane were measured to  locate $T_{m}$ of each sample. At temperatures above $T_{m}$, the transverse magnetization relaxation has a gaussian decay principally associated with dynamics of the dipolar local  field from near-neighbor copper discussed by Bachman {\it et al.}~\cite{Bachman1998}. Upon entry into the vortex solid state, however, a new relaxation mechanism onsets due to thermal fluctuations of vortices that are formed into a vortex lattice. Their lorentzian spectral density produces an exponential component to the time dependence of the nuclear magnetization relaxation~\cite{Bachman1998}.  Spin-spin relaxation rates from three different nuclei, i.e., planar Cu, planar oxygen, O(1), and the apical oxygen, O(2), in our UD87 sample are shown in Fig. \ref{fig1} in a magnetic field $H=6.4$\,T. The onset of slow localized vortex field fluctuations is marked by a peak in $T_{2}^{-1}$ near $T=40$\,K,  independent of nucleus or of field orientation. 

\begin{figure}[!ht]
\centerline{\includegraphics[width=0.5\textwidth]{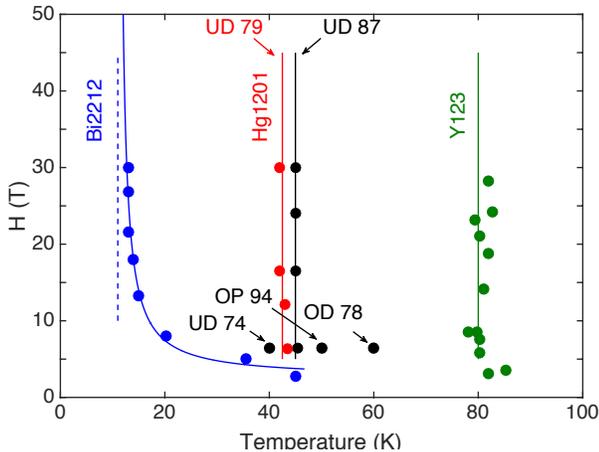}}
\caption{The vortex phase diagram for a number of cuprate superconductors shows sharp upward curvature with a field independent behavior for formation of the vortex lattice.  This was theoretically predicted to occur at sufficiently high field  by Glazman and Koshelev~\cite{Glazman1991} in the  two dimensional limit, applicable to highly anisotropic HTS, and was experimentally observed in Bi2212 by Chen {\it et al.}~\cite{Chen2007}.  The vertical dashed line indicates the high-field, two-dimensional limit of the melting transition.  However, similar behavior was observed for Y123 (Bachman {\it et al.}~\cite{Bachman1998}) and in Hg1201 (this work), indicated by solid vertical lines, but cannot be understood in the same context (see text).} 
\label{fig2}
\end{figure}

For strongly anisotropic superconductors the magnetic field-temperature vortex melting phase diagram is predicted to have a limiting behavior at high applied magnetic field. This behavior is characterized by a field-independent melting transition temperature provided that the applied field is larger than a crossover field that varies inversely with the square of the mass anisotropy, $\gamma$. In this high field limit electromagnetic interaction between vortices in the CuO$_{2}$ plane dominates inter-planar Josephson coupling resulting in quasi two-dimensional vortex dynamics of pancake vortices. As the field is increased $T_{m}$ asymptotically approaches the following 2D limit,

\begin{equation}
\label{eq1}
T_{m}^{2D} = A\frac{\Phi_0^2d}{8\pi\sqrt{3}k_{b}(4\pi\lambda_{ab}(0))^2}.
\end{equation}

\noindent
The flux quantum is, $\phi_0$, and the interlayer separation is  $\it{d}$, which is 0.95\,nm for Hg1201.  The zero temperature limit of the penetration depth for currents in the $ab$-plane is denoted by $\lambda_{ab}$(0). The constant, $\it{A}$ $\sim0.61$, depends on the Lindemann criterion for melting and can be numerically computed~\cite{Brandt1989,Brandt1989-2,Vinokur1998,Koshelev2007,
 Chen2007}.  Most importantly the observed 2D limit found for optimally doped Bi2212, $T_{m}^{2D}=12$\,K corresponds to a penetration depth $\lambda_{ab}(0)= 220$\,nm  consistent with independent measurements~\cite{Kogan1993, Prozorov2000} falling in their range from 180 to 270\,nm thereby providing strong affirmation for the theory~\cite{Glazman1991}. For Hg1201 UD79 and UD87 we find no apparent field dependence of $T_{m}$ from $H = 6.4$ to 30.0 T, suggesting the crystals are already in the 2D lattice limit.  However, this is not the case.  The observed  high field limit for melting is 42\,K for UD79 and 45\,K for UD87, shown in Fig.\ref{fig2}.  According to the 2D melting theory this would correspond to  penetration depths of 76\,nm and 74\,nm respectively, considerably at odds with the previous reports  171\,nm \cite{Panagopoulos1997} and with our measurements $\sim190$\,nm discussed in section IV.  The same discrepancy exists for optimally doped Y123 where the 2D limit would imply a penetration depth of $\sim61$\,nm at odds with the reported value of $\sim150$\,nm~\cite{Zimmermann1995}.  We infer that the field independence of the melting, defined by abrupt changes in vortex dynamical time scales and their spectral density~\cite{Bachman1998} must have a different explanation.  

\begin{figure}[!ht]
\centerline{\includegraphics[width=0.5\textwidth]{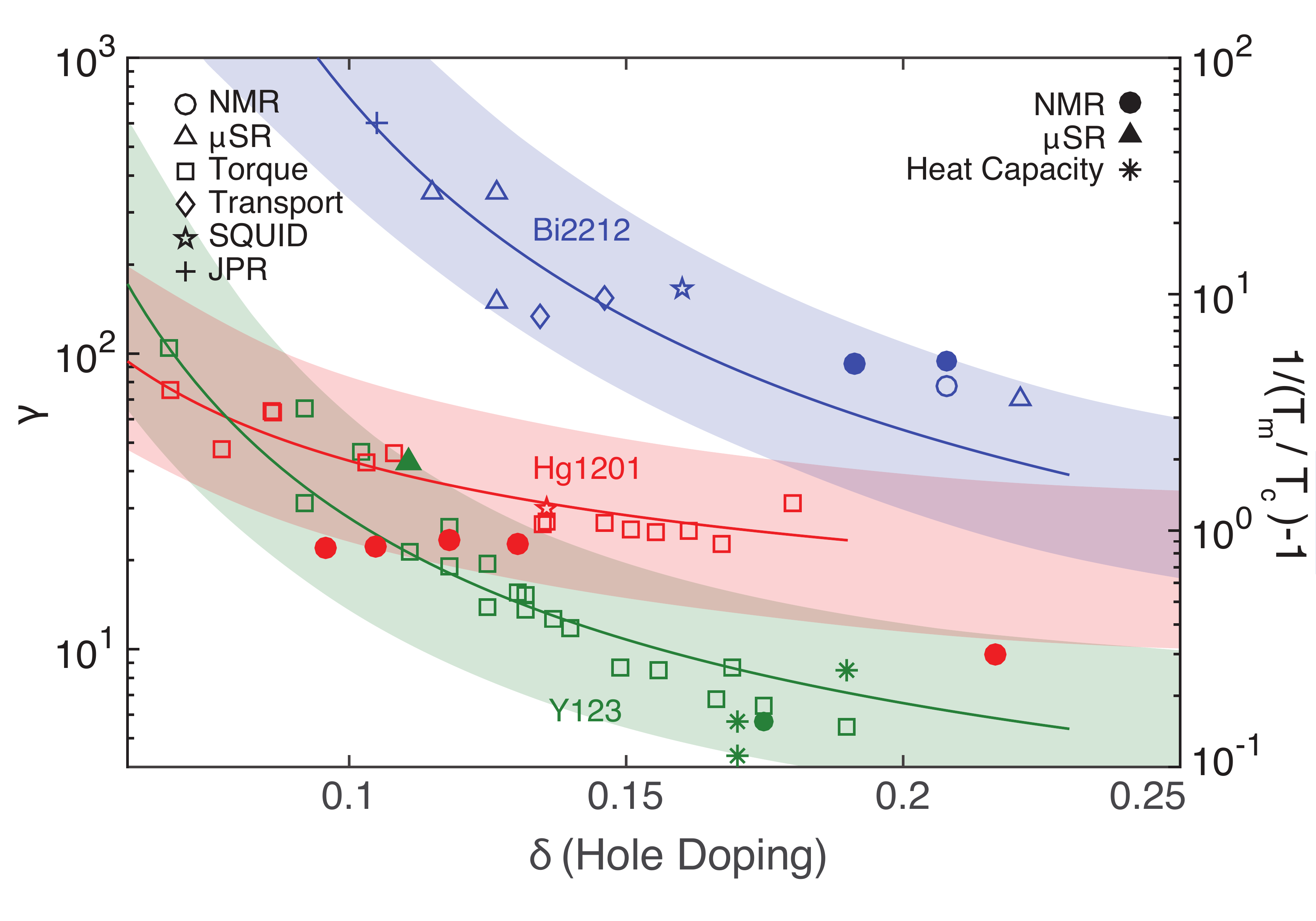}}
\caption{The mass anisotropy for three HTS compounds is shown as  function of doping compared with the temperature for melting, $T_m$.  Data from NMR \cite{Chen2007Thesis,Chen2007}, $\mu$SR \cite{Bernhard1995}, torque \cite{Janossy1992,Chien1994,Xia2012,Hofer1998}, transport (open symbols) \cite{Lee1993,Lee1995,Jacobs1995},  SQUID (star) \cite{Giannini2008,LeBras1996}, and Josephson plasma resonance (cross) measurements \cite{Colson2003} indicate that Bi2212  is much more anisotropic than Hg1201 or Y123~\cite{Giannini2008,Colson2003}. 
The behavior for Bi2212 is consistent with the theory for 2D melting~\cite{Glazman1991} but is not appropriate for Hg1201 or Y123 as discussed in the text.  The melting transitions determined at low field from $\mu$SR \cite{Sonier2000}, NMR \cite{Bachman1998,Chen2007,Chen2007Thesis}, and heat capacity \cite{Roulin1998,Schilling1996,Schilling2002} are also shown for comparison.}
\label{fig3}
\end{figure}

The most obvious difference in the comparison of  Bi2212 with the other two compounds, Hg1201 and Y123, is in their much smaller mass anisotropy.  This point is made explicit in Fig.~\ref{fig3} where the temperature difference between the vortex melting transition and the superconducting transition is compared with the mass anisotropy as a function of doping  over a wide range taken from the literature~\cite{Janossy1992,Chien1994,Schilling1996,Roulin1998,Bachman1998,Sonier2000,Schilling2002,Lee1995,Lee1993,Jacobs1995,Bernhard1995,Colson2003,Chen2007Thesis,Chen2007,Giannini2008,LeBras1996,Hofer1998,Xia2012}.  We emphasize that the melting transitions observed from the NMR relaxation measurement  do not identify a thermodynamic phase transition. The change in the vortex dynamic time scale could well be an indication of a crossover in the behaviors of vortices.  Nonetheless, generality of the field independent $T_{m}$ observed in a number of different HTS compounds is compelling and appears to correlate well with the mass anisotropy.

\section{Vortex Lattice Structure}

\begin{figure}[!ht]
\centerline{\includegraphics[width=0.5\textwidth]{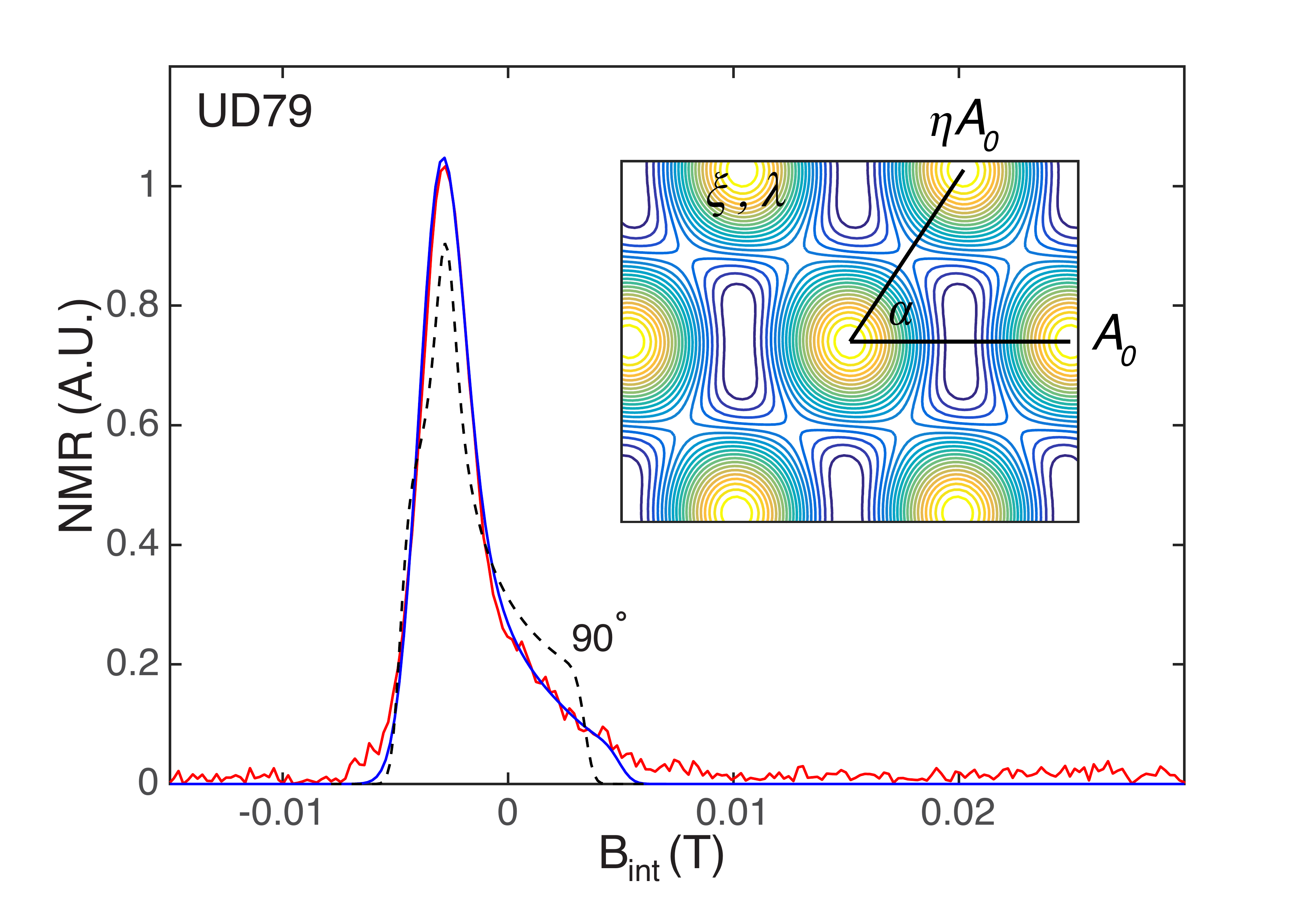}}
\caption{The vortex magnetic field distribution versus internal magnetic field, $B_{int}$, relative to the applied field. The NMR spectrum shown in the figure was obtained for the O(2)(3/2,1/2) transition at 35\,K in the applied field $H = 16.5$\,T compared with the calculated spectrum determined using Brandt's algorithm, optimized to fit the experimental spectrum using the Levenberg-Marquardt method.  A non-vortex contribution to the spectrum was measured above the vortex melting transition temperature for the same NMR quadrupolar satellite and at the same field which was then convolved with a raw spectrum to obtain the vortex field distribution.  A fit constrained to be for a tetragonal lattice ($\alpha$ = 90$^{\circ}$) is shown  for comparison as a black dashed line. The inset figure displays the magnetic field contours for the set of four VL related parameters in the calculation.} 
\label{fig4}
\end{figure}

Upon cooling into the vortex solid state, a new NMR lineshape emerges from the normal state spectrum that is a convolution with the vortex spectrum commonly known as the Redfield pattern, Fig.~\ref{fig4}, characterized by a long high frequency tail reflecting local magnetic fields near or inside the cores of vortices. The Redfield pattern is simply proportional to the probability distribution for local magnetic fields that is associated with the vortex state probed by the nuclei in the crystal.  It can be used to construct the vortex lattice structure making NMR a powerful complimentary tool to other experimental techniques such as $\mu$SR and small angle neutron scattering (SANS) ~\cite{Eskildsen2002}.

Since the spatial information of the VL is represented by an average encoded in a single NMR spectrum, one needs to decode it to reconstruct the lattice structure. This is done by fitting a spectrum to a probability density distribution of magnetic fields from an ideal VL configuration that minimizes the Ginzburg-Landau (GL) free energy~\cite{Mitrovic2001}.

We have used a fast convergent iterative method developed by Brandt \cite{Brandt1997} to find a GL solution based on the following parameters that define the VL: the coherence length, $\xi$; the penetration depth, $\lambda$; the VL lattice constant, A$_{0}$; and the lattice obliqueness angle, $\alpha$. The lattice constant anisotropy, $\eta$,  defined in  Fig. \ref{fig3}, can be calculated from the vortex density equation assuming single flux quantization,

\begin{equation}
\label{eq4}
\eta = \frac{\Phi_{0}}{H_{0}A_{0}^2\sin\,\alpha}.
\end{equation}

\begin{figure}[!ht]
\centerline{\includegraphics[width=0.5\textwidth]{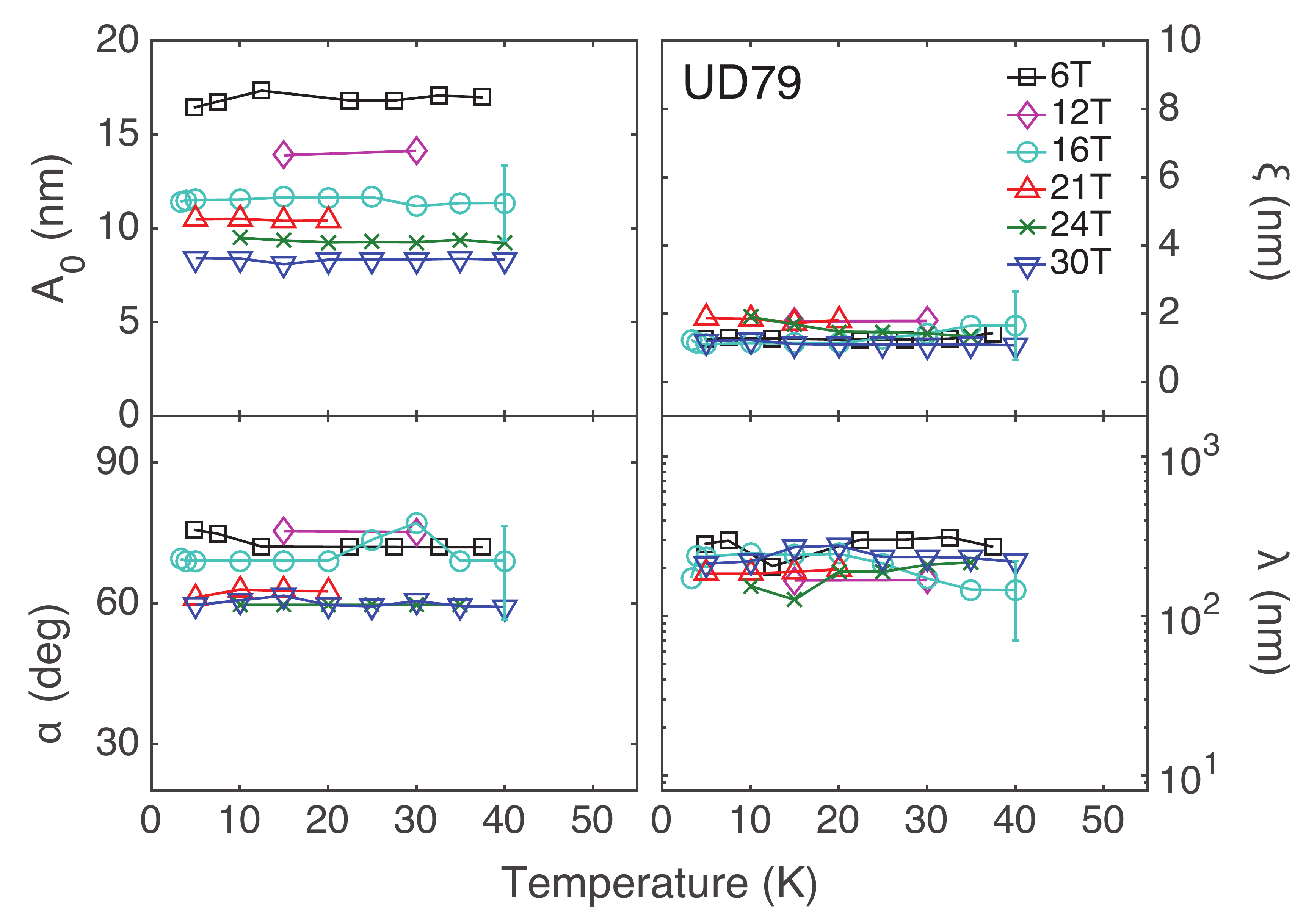}}
\caption{Vortex lattice properties in Hg1201 obtained from lineshape analysis at various fields as a function of temperature.  These are the VL lattice constant, A$_{0}$; the lattice obliqueness angle, $\alpha$; the coherence length, $\xi$; and the penetration depth, $\lambda$. The statistical error  representative of each data set is shown as a single error bar.  Within  statistical error there is no discernible field and temperature dependence of the penetration depth and coherence length. The low field data from $H$ = 6.4\,T was obtained from vortex fitting of the central transition, (1/2,-1/2), spectra from the planar $^{63}$Cu site (6\,T). The remainder of the data was obtained from planar $^{17}$O at the (3/2, 1/2) transition.} 
\label{fig5}
\end{figure}

\begin{figure}[!ht]
\centerline{\includegraphics[width=0.5\textwidth]{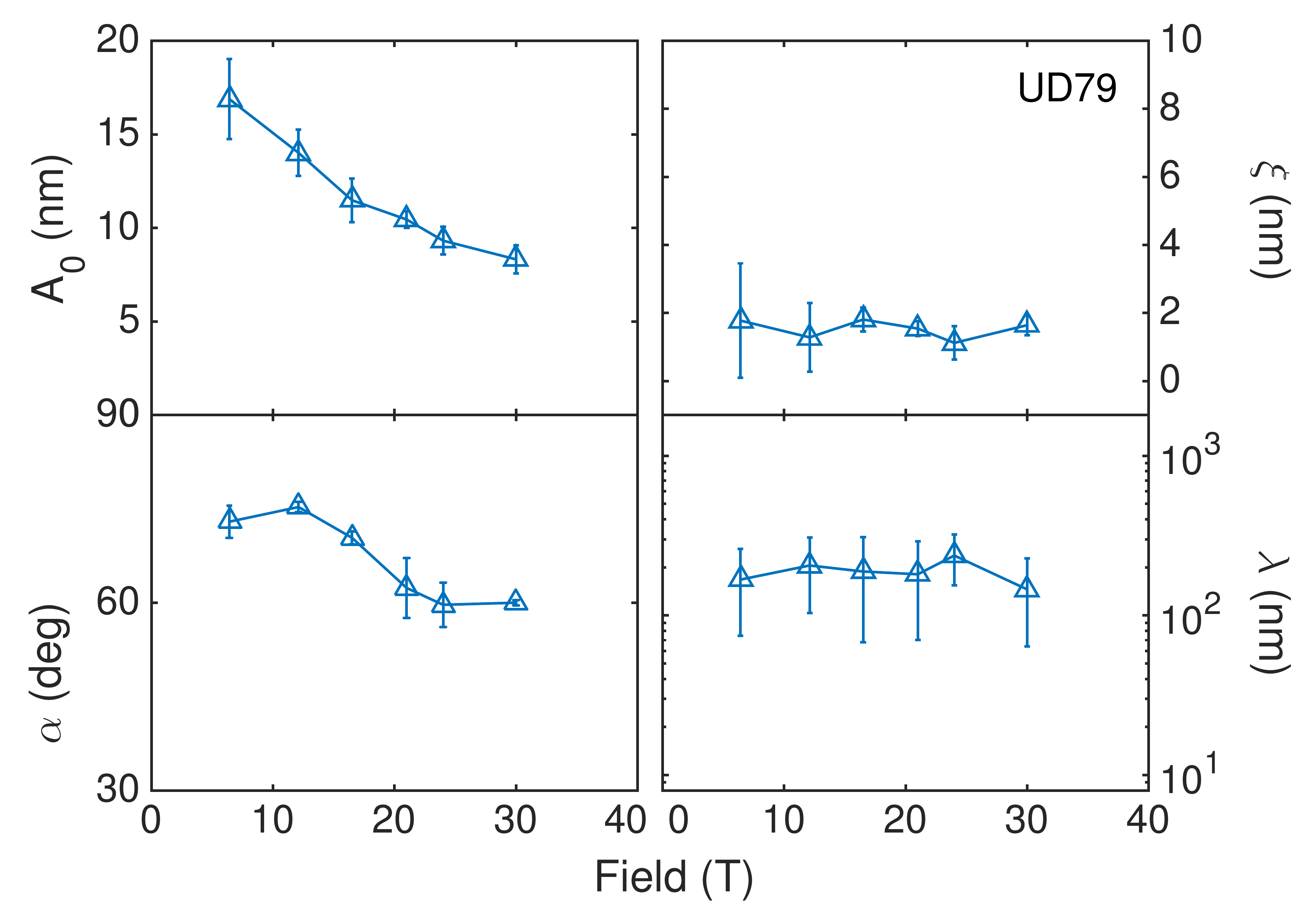}}
\caption{The result of the vortex analysis for the vortex structure parameters averaged over temperature at each field as a function of field. The decreasing trend in the VL spacing, $A_{0}$, mostly reflects the increasing vortex density as the external field is increased. It is also affected by the VL angle as shown in Eq.  \ref{eq4}. The lattice obliqueness, $\alpha$, indicates a  vortex structural transition from an orthorhombic to a triangular lattice near 15\,T.} 
\label{fig6}
\end{figure}

Two parameters, $\alpha$ and $\eta$,  represent an angle between two vectors defining a vortex lattice and length anisotropy ratio between the lattice vectors, respectively. For example, a triangular VL is given by ($\alpha$, $\eta$) = (60$^\circ$,\,1) or an elongated rectangular lattice by (90$^\circ$,\,$\neq$1). Typically, about 20 iterations are needed to obtain a single GL solution using  Brandt's algorithm. Once a solution was found, a next set of vortex parameters was estimated based on the Levenberg-Marquardt optimization criteria and this entire process was iterated for convergence. This fitting process is completed typically after 60-70 iterations and a best spectrum with the result of the fit is shown in Fig. \ref{fig4}, with the optimized parameter sets as a function of temperature and magnetic field shown in Fig.~\ref{fig5} and Fig.~\ref{fig6}. 

The results from our analysis show that $\lambda$ and $\xi$ are field and temperature independent but the structure of the VL inferred from $\alpha$ and $\eta$  suggests a field-induced structural transition for $H$ between 10\,T and  20\,T, from an oblique lattice with $\alpha$ = 73$\pm$7$^{\circ}$ to a triangular lattice with $\alpha$ = 60$\pm$5$^{\circ}$, Fig.~\ref{fig6}.

In Y123, a SANS experiment  revealed a low field first-order structural transition in the material, from a triangular lattice below 6.5\,T to an oblique lattice at 10.8\,T\cite{Keimer1994,Maggio-Aprile1995,White2009}. This  transition reflects the four-fold symmetry $\it{d_{x^2-y^2}}$ of the superconducting order parameter \cite{Ichioka1999} based on a GL calculation which shows a lower free energy with the orthorhombic oblique structure. If there is such a transition in Hg1201 it must occur at lower magnetic fields than 6.4\,T.

There is no other experimental evidence for a high field vortex transition to a triangular structure. In theoretical work with more detailed treatment of  the $d$-wave order parameter in the vortex state,~\cite{Berlinsky1995, Al-Saidi2003} it was shown that anisotropic $d$-wave superconductors can undergo a secondary transition due to a mixed pairing state in high fields near the upper critical field ($H \sim0.8 H_{c2}$).  Field dependence of the free energies of multiple vortex configurations was also calculated by Ichioka {\it et al.}~\cite{Ichioka1999} providing an indication of a possible structural transition at high fields in which a triangular configuration is favored over other shapes due to a lower free energy. 

\section{Conclusion} 
The single layer Hg1201 cuprate compound offers an  attractive  platform to investigate vortex structure and dynamics using $^{17}$O NMR. This material is available in single crystals that can be tuned from underdoped to overdoped where the dopant oxygen is located in the HgO plane, far from the apical and planar oxygen sites in the crystal. The NMR spectra are narrow, and in the case of the apical quadrupolar satellites,  immune from overlap with other resonances,  and relatively immune from hyperfine coupling to the electronic structure in the CuO$_2$ plane and electronic disorder and charge order modulations.  Using  the apical oxygen quadrupolar satellite resonance we have determined a magnetic field independent  melting behavior of vortices up to high magnetic field, 30\,T, and show that the vortex interactions deviate substantially from two dimensionality, different from Bi2212 but similar to Y123 compounds.  Our analysis of the vortex structure in Hg1201 indicates a vortex lattice transition from oblique at low magnetic fields, $\sim6$\,T,  to triangular at high fields above 15\,T.
\newline \newline

\section{Acknowledgments}
We acknowledge contributions from Andrew Mounce, discussions with  Vesna Mitrovic,  and we thank Martin Greven for supplying the unprocessed Hg1201 crystals.   Research was supported by the U.S. Department of Energy, Office of Basic Energy Sciences, Division of Materials Sciences and Engineering under Awards DE-FG02-05ER46248 (Northwestern University).  M.K.C. is supported by funds from the US Department of Energy BES grant no. LANLF100.  A portion of this work was performed at the National High Magnetic Field Laboratory, which is supported by National Science Foundation Cooperative Agreement No. DMR-1157490 and the State of Florida.

\bibliography{VortexTransitionInHg1201}

\end{document}